\documentclass[twocolumn,amsmath,amssymb,superscriptaddress,nobibnotes,aps,tightenlines,prl]{revtex4}
\usepackage{graphicx}
\usepackage{epsfig}
\usepackage{amsmath}
\usepackage{bm}
\usepackage{setspace}
\usepackage{color}
\usepackage{gensymb}

\begin{document}

\title{Full-dimensional theory of pair-correlated HNCO photofragmentation}

\author{L. Bonnet$^*$}
\affiliation{Institut des Sciences Mol\'eculaires, Universit\'e de Bordeaux, CNRS, UMR 5255, 33405, Talence, France}
\thanks{
 Corresponding authors. Email: claude-laurent.bonnet@u-bordeaux.fr; hochlaf@univ-mlv.fr; 
}

\author{R. Linguerri}
\affiliation{Laboratoire Mod\'elisation et Simulation Multi Echelle, Universit\'e Paris-Est, UMR 8208 CNRS, 5 Bd Descartes,  77454 Marne La Vall\'ee, France}

\author{M. Hochlaf$^*$}
\affiliation{Laboratoire Mod\'elisation et Simulation Multi Echelle, Universit\'e Paris-Est, UMR 8208 CNRS, 5 Bd Descartes,  77454 Marne La Vall\'ee, France}

\author{O. Yazidi}
\affiliation{Laboratoire de Spectroscopie Atomique, Mol\'eculaire et Applications, LR01ES09, Facult\'e des Sciences de Tunis, Universit\'e de Tunis El Manar, 2092 Tunis, Tunisia}

\author{P. Halvick}
\affiliation{Institut des Sciences Mol\'eculaires, Universit\'e de Bordeaux, CNRS, UMR 5255, 33405, Talence, France}

\author{J. S. Francisco}

\affiliation{Department of Chemistry, University of Nebraska-Lincoln, 433 Hamilton Hall, Lincoln, Nebraska 68588-0304, USA}

\date{\today}

\begin{abstract}
\noindent 
Full-dimensional semiclassical dynamical calculations are reported for the photofragmentation of isocyanic acid in the S$_1$ state. 
These calculations, performed for the first time, allow to closely reproduce the key features of high-resolution imaging measurements 
at photolysis wavelengths of 201 and 210 nm while providing insight into the underlying dissociation mechanism.
\end{abstract}

\maketitle
Isocyanic acid, HNCO, is the simplest molecule that contains H, C, N and O atoms, which are essential 
for life and major constituents of chemical species in chemistry and biology. Since its discovery in 1830 \cite{Liebig1830}, 
HNCO has been identified in diverse media.   
For instance, this molecule is found in over sixty galactic sources and in nine external galaxies 
\cite{Marcelino2010,Snyder1972,Buhl1972,Zinchenko2000,Nguyen1991,Jackson1984}. It is relatively
 highly abundant and it is suggested to be strongly involved in prebiotic chemistry occurring there. 
Especially, it is a major component of the photochemistry of building blocks of life (DNA bases and amino acids). Isocyanic acid has been 
identified in the atmosphere \cite{Perry1986,Miller1991} where HNCO sources are from both 
anthropogenic and biomass burning, and also released from combustion (e.g. fuels, cigarettes \cite{Bowman1992,Roberts2011}). 
It is known to be toxic and easily dissolves in water (in lungs) posing obvious health risks such as 
the development of atherosclerosis, cataracts and rheumatoid arthritis \cite{Roberts2011,Young2012} and
 inflammations that can lead to cardiovascular disease. 
To-date, the health effects of isocyanic acid, its role in the atmosphere and its photochemistry are not fully known nor understood.

The photochemistry of HNCO under UV radiations or in combustion sources is complex. Besides the importance of isocyanic acid in diverse fields 
(see above), its unimolecular decomposition processes represent benchmarks for tetratomic molecules evolution upon electronic excitation. 
Adiabatically, the lowest HNCO singlet electronic states cannot lead to the formation of NH (X$^3\Sigma^-$) + CO(X$^1\Sigma^+$) products. 
Instead, the observed formation results from electronically excited NH(a$^1\Delta$) photofragment (with CO(X$^1\Sigma^+$)). 
Previous studies showed that around 200 \textcolor{black}{and 210} nm wavelengths, \textcolor{black}{considered further in this work,} 
HNCO mainly photodissociates in the first excited
 singlet state S$_1$ \cite{Zyrianov,Klo,Laine2001,Spiglanin1987,Spiglanin1987_2,Yi1996,Brownsword1996,Brownsword1996_2,Brownsword1997,Brownsword1999}.
 \textcolor{black}{For lower optical excitations,} internal conversion to the S$_0$ ground state ($^1$A$^\prime$ (1 $^1$A)) through seams of conical 
intersections with the S$_1$ state ($^1$A$^{\prime\prime}$ (2 $^1$A)) leads to the photofragments, H + NCO 
\cite{Yarkony2016,Yarkony2001}, by spin-allowed 
unimolecular decay, \textcolor{black}{or} to NH(X$^3\Sigma^-$) + CO(X$^1\Sigma^+$), after intersystem crossing and decomposition on the lowest 
triplet state potential.

Up-to-date experimental investigations of HNCO photodissociation have been performed by Yang and co-workers \cite{Wang2007,PY}.
In 2007, they \cite{Wang2007} \textcolor{black}{measured the pair-correlated angular and kinetic energy distributions (KEDs) 
for the NH(a$^1\Delta$) + CO(X$^1\Sigma^+$) 
channel using velocity map imaging (VMI), where CO was state-selectively detected by (2+1) resonance-enhanced multi-photon ionization (REMPI).
They proved the predominance of the formation of the previous products following photon excitation at 210 nm \cite{Wang2007}, with
co-product NH(a$^1\Delta) \equiv\;^1$NH in the vibrational ground state. Rotational state distributions of $^1$NH were deduced from the KEDs 
while state-resolved anisotropy parameters (APs) were deduced from angular distributions. They showed that the produced CO fragments are 
rotationally hot (with population up to $j_{CO} = 50$) and that $^1$NH \textcolor{black}{co-products} are, however, rotationally cold.
Moreover, APs are negative, thus indicating that the photofragmentation of HNCO at 210 nm is mainly direct.
More recently, they \cite{PY} repeated the analogous experiment at 201 nm, where $^1$NH was state-selectively detected via REMPI,
and arrived at the following conclusions: the vibrational state populations (VSPs) of CO are inverted for nearly all the rotational states 
of $^1$NH in its vibrational ground state; APs are negative, showing once again the direct nature of the dissociation; rotational state distributions 
of CO in the vibrational ground state are bimodal. Zhang \emph{et al.}~\cite{PY} interpreted the bimodality as resulting from the involvement 
of two HNCO isomers ($cis$ and $trans$) in the fragmentation dynamics. The goal of this letter is to theoretically reproduce this correlated 
information on HNCO photodissociation around 200 and 210 nm and provide insight into its underlying dynamics,  in particular, validating 
 the speculated mechanism.}

\begin{figure}
\advance\leftskip-1cm
\includegraphics[width=50mm]{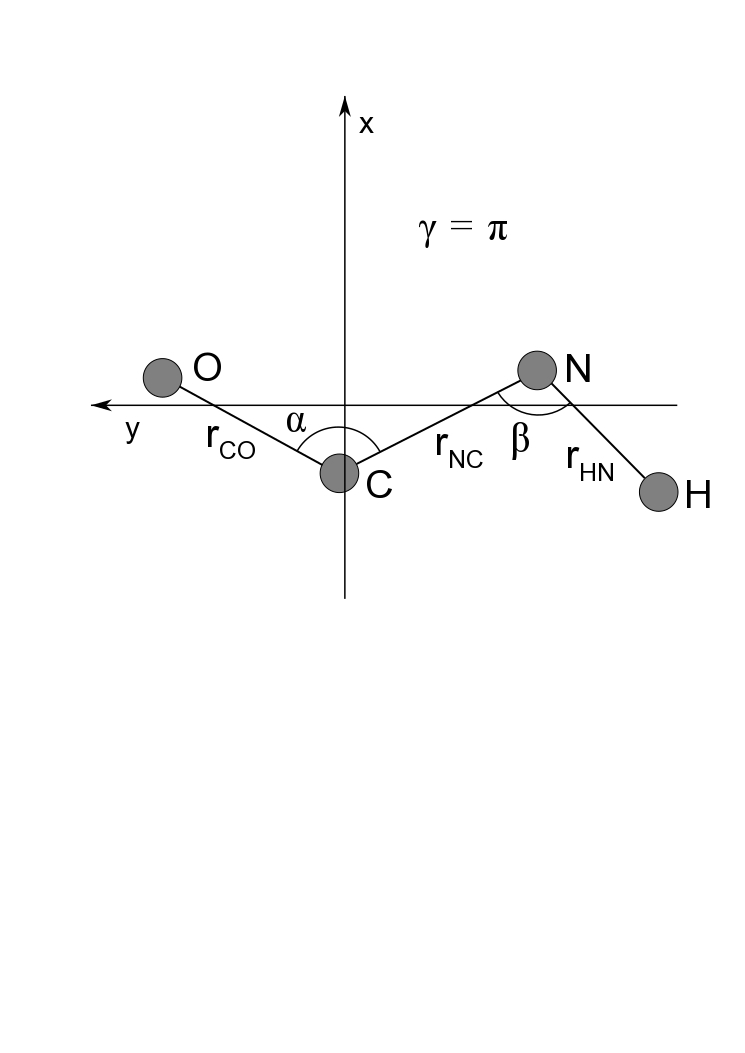}
\caption{
Internal coordinates used for the description of the HNCO PES. $\gamma$ is the dihedral angle around the NC bond
($\gamma=\pi$ for the $trans$ configuration). We also give the body-fixed frame convention.}
\label{figure1}
\end{figure}

The exact quantum treatment of the photodissociation dynamics of tetratomic molecular systems is at the capacity limit of today's theoretical 
and numerical techniques \cite{Xie2014}. This is particularly true for the title process, as four different atoms are involved and no 
symmetry can be used to converge calculations by means of reasonable basis sizes (not to mention the significant mass of the three second raw 
atoms that is a complicating factor). Therefore, we used the newly developed 
classical trajectory method (CTM) in a quantum spirit \cite{Part3}. This semiclassical approach 
proved to lead to nearly quantitative predictions for several bimolecular, unimolecular and vibrational predissociation processes 
\cite{Part3, Chinese-C+H2}. As for photodissociations, CTM in a quantum spirit 
involves the three following steps: (i) selecting the set of initial conditions of trajectories from the Wigner distribution associated with 
the internal state of HNCO before the optical excitation \cite{Goursaud, Schinke}; (ii) running trajectories from these initial 
conditions up to the separated products \cite{Schinke}; and (iii) deducing from the final conditions the pair-correlated KEDs, VSPs and APs by 
applying Bohr quantization rule \cite{Bohr} to product vibrational motions \cite{Part3}. This basic rule of quantum mechanics states in a 
language suited to the present problem that one should only take into account those classical trajectories reaching the products with integer 
vibrational actions. This has a significant influence on the final results when only a small number of vibrational states are available to 
the products, as is the case for the fragmentation of HNCO at 201 and 210 nm. 
On the other hand, between ten and several tenths of rotational states are available to the products for this process, so rotational motions 
need not be pseudo-quantized. Strict application of Bohr's condition of quantization is, however, too restrictive from a numerical point of 
view and, instead, trajectories are usually assigned Gaussian statistical weights such that the closer the vibrational actions to integer values, 
the larger the weights. This procedure is known as Gaussian binning (GB) \cite{Part3} and CTM in a quantum spirit will thus be
called GB-CTM. Within the framework of bimolecular collisions, GB has been shown to be 
consistent \cite{Part3} with the semiclassical theory of molecular collisions pioneered by Miller and Marcus \cite{Miller,Marcus}.
More technical details on the GB-CTM calculations presented further below can be found in the Supplemental Material. 
The coordinate system used in this work is represented in Fig.~\ref{figure1}.

None of the previous theoretical treatments of the title reaction were performed in full-dimensionality \cite{Klo,Wang2007}.
Instead, HNCO was kept planar throughout its fragmentation, being thus reduced to a five-dimensional (5D) system. For the sake of realism, however, 
out-of-plane motions should be taken into account. About two decades ago, Klossika and 
Schinke \cite{Klo} built an $S_1$ planar potential energy surface (PES) fitted to high-level \emph{ab-initio} calculations (KS-PES). 
We thus took advantage of the availability of this high-quality KS-PES to which we added a contribution depending mainly on $r_{NC}$ and the dihedral 
angle $\gamma$, fitted to equation of motion coupled-cluster calculations. 
This modification is detailed in the Supplemental Material. Panel A in Fig.~\ref{figure2} displays the bi-dimensional cut of the resulting 6D-PES 
along $r_{NC}$ and the $\alpha$ angle, together with two typical trajectories discussed later below. Panel B in Fig.~\ref{figure2} shows a similar 
cut along $r_{NC}$ and the $\gamma$ torsional angle (for both figures, the remaining coordinates were kept at the 
values corresponding to the trans minimum in the $S_1$ state, given in Table V of ref.~\cite{Klo}). It is worth noting that r$_{NC}$ and 
$\gamma$ are strongly coupled to each other, which necessarily modifies the dynamics as compared to those obtained in a planar description. 
\begin{figure}
\advance\leftskip-0.5cm
\includegraphics[width=80mm]{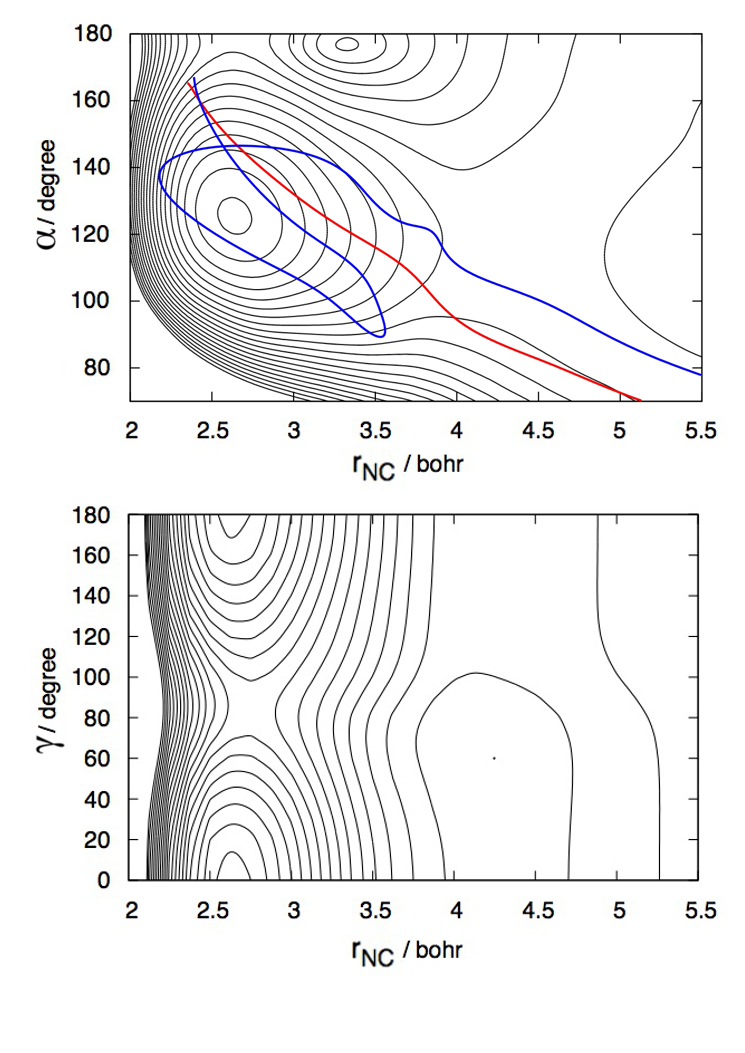}
\caption{A) 2D cut of the S$_1$ potential along $r_{NC}$ and the NCO $\alpha$ angle. The red and blue lines are two typical 
trajectories (see text). B) Similar cut along $r_{NC}$ and the dihedral $\gamma$ angle.}
\label{figure2}
\end{figure}

\begin{figure}
\advance\leftskip-0.5cm
\includegraphics[width=90mm]{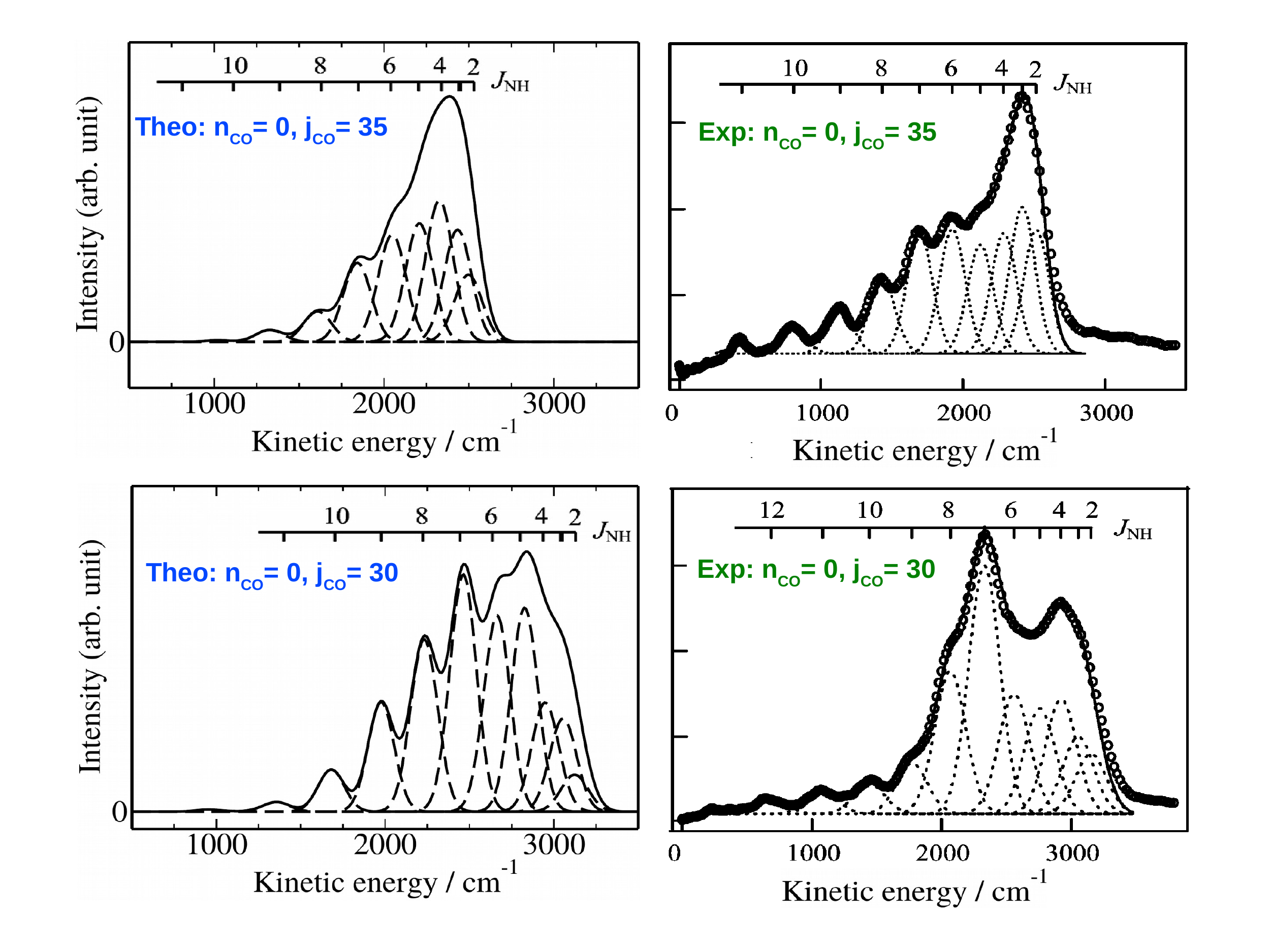}
\caption{GB-CTM (left) and experimental (right) \cite{Wang2007} pair-correlated kinetic energy distributions at 210 nm. The Dirac combs 
locate the translational energy consistent with $(n_{CO},j_{CO})$ and $(n_{NH}=0,j_{NH})$ (assuming a perfect energy-resolved experiment).}
\label{figure3}
\end{figure}

The GB-CTM KEDs at 210 nm are compared with their experimental counterpart in Fig.~\ref{figure3} (solid lines). These distributions are deduced 
from slicing images of CO in the vibrotational states ($n_{CO},j_{CO}$) = (0,30) and (0,35) \cite{Wang2007}. The energy available 
to the products is such that $^1$NH can only be in the vibrational ground state. The contributions assigned to the $j_{NH}$ rotational levels are 
also shown (dashed curves). As a matter of fact, the main features of the experimental KEDs (threshold and cut-off kinetic energies, positions of the 
rotationally-resolved peaks, overall shape) are well reproduced by our theoretical treatment. In particular, both sets of data show the 
production of rotationally cold $^1$NH molecules, with $j_{NH}$ smaller than $\sim$ 10.  

\begin{figure}
\advance\leftskip-0.5cm
\includegraphics[width=90mm]{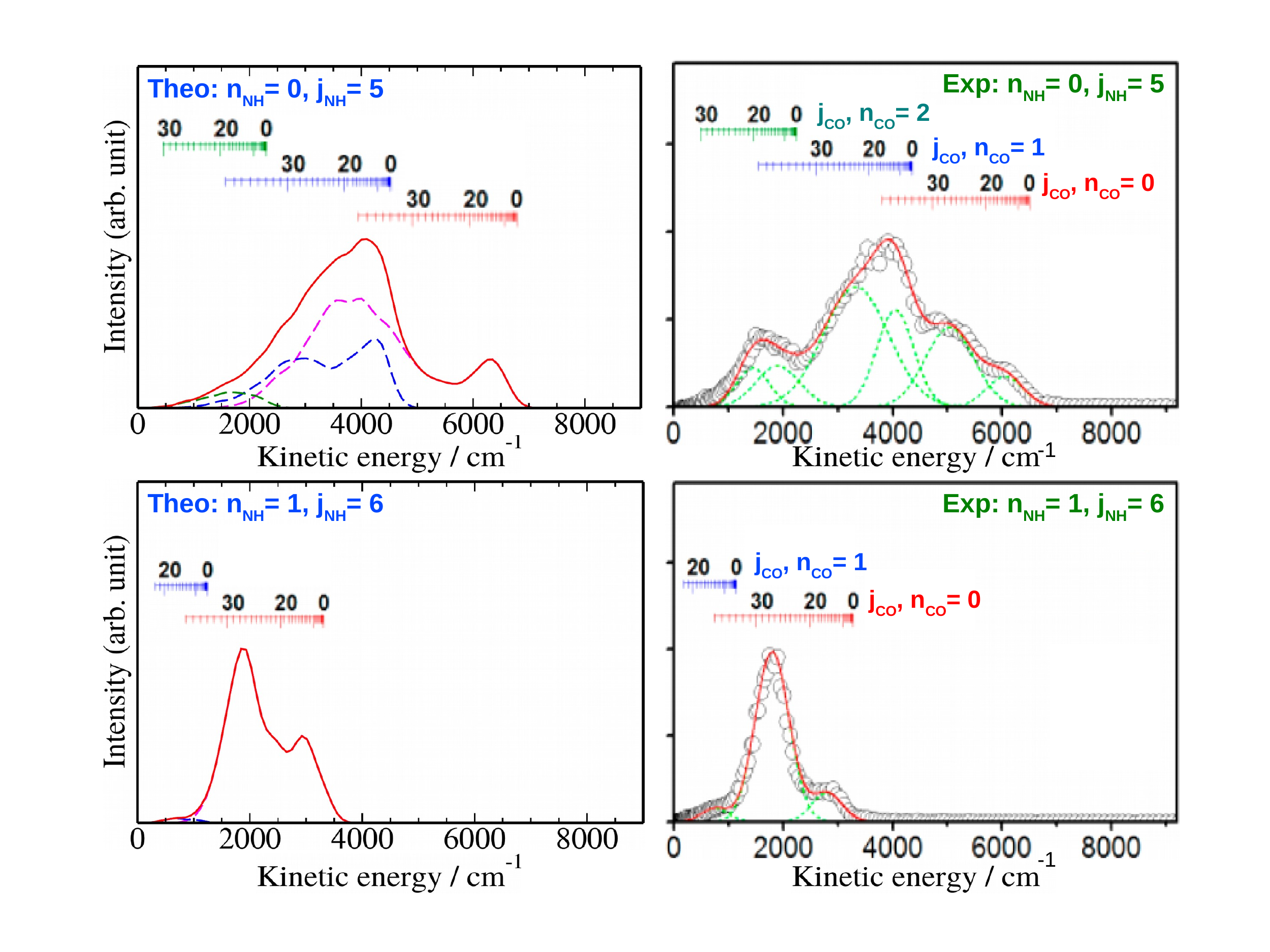}
\caption{GB-CTM (left) and experimental (right) \cite{PY} pair-correlated kinetic energy distributions at 201 nm. Each CO vibrational state
is assigned a Dirac comb to locate the translational energy consistent with $(n_{NH},j_{NH})$ and $(n_{CO},j_{CO})$. The correspondence 
between combs and $n_{CO}$ is shown in the right panels.}
 \label{figure4}
\end{figure}

Fig.~\ref{figure4} shows similar comparison at 201 nm. The displayed experimental KEDs were obtained from slicing images of $^1$NH in the 
vibrotational states ($n_{NH},j_{NH}$) = (0,5) and (1,6) \cite{PY} (see Figs.~S1-S3 for more $^1$NH states). Again, the overall agreement 
between experiment and theory (solid curves) is very satisfying (the best agreement being achieved for (0,9), (1,6), (1,8)). 
The bimodality of the rotational state distributions of CO in the vibrational ground state, seen in the right-lower panel of 
Fig.~\ref{figure4}, is clearly reproduced in the left-lower panel. The production of rotationally hot CO, discussed further below, 
is striking in both the experimental and theoretical results. This is also the case for the (1,4) and (1,8) states (see Fig.~S3).
We wish to emphasize that the quality of the accord for both wave lenghts would be lost without respecting Bohr quantization, 
as recently illustrated in Ref.~\cite{BCE}. The dashed curves in the left panels are the CO vibrational state contributions to the total KED. 
Interestingly, most of them (see also Figs.~S1-S3) are bimodal to some extent. 
The areas below the dashed curves are proportional to the populations of their corresponding $n_{CO}$ values, thus leading 
to the $^1$NH state-resolved VSD of CO after \emph{ad-hoc} renormalization to unity (see Fig.~S4). One thus sees 
from Figs.~\ref{figure4} and S1-S3 that theory does not predict any vibrational inversion at 201 nm, 
whatever the $^1$NH state probed. This is in contrast with the conclusions of Yang and co-workers \cite{PY} who found that the VSDs of CO peak at 1 
for $n_{NH}$=0 and nearly all the values of $j_{NH}$. However, these authors recognize that one cannot avoid any ambiguity in the CO vibrational 
assignment of a given pair-correlated KED when its $n_{CO}$-resolved components overlap \cite{PY}.

The theoretical APs (between -0.95 and -0.8 at both 210 and 201 nm; see Fig.~S4) are slightly lower than the measured ones 
(between -0.8 and -0.5). Nevertheless, 
our approach corroborates the experimental finding that the fragmentation proceeds through a fast direct impulsive mechanism. This 
is also consistent with the short theoretical lifetime of HNCO between its optical excitation and its fragmentation (305 and 175 fs at 210 
and 201 nm, respectively). We discuss in the Supplemental Material how relaxing two constraints ($\bm{\mu} \parallel z$; internal conversion 
to S$_0$ ignored) in our calculations should improve the predictions.

Finally, in order to check the validity of the supposed link between (i) bimodality of rotational state distributions and (ii) involvement 
of two $cis$ and $trans$ HNCO isomers in the fragmentation dynamics \cite{PY}, we have calculated the distribution of the dihedral angle 
$\gamma$ at $r_{NC} = 4.1$ bohr, distance nearly defining both transition states (TSs) (see Fig.~S5). For the vast majority 
of the trajectories ($\sim 95\%$), $\gamma$ is larger than 90\degree, which separates the basins of attraction of the $cis$ and $trans$ HNCO isomers 
and their associated TSs. $\gamma$ being equal to 180\degree\;at the $trans$ TS, the $cis$ TS crossing is clearly a minor event. Note that at the 
Franck-Condon (FC) geometry ($\alpha=172$\degree), the system lies within the $trans$ basin. Though selecting initial 
conditions according to the Wigner distribution necessarily generates many trajectories with an initial motion directed towards the $cis$ basin, 
the forces towards the $trans$ basin result 
sufficiently strong to redirect the trajectories towards it. Consequently, the bimodality observed theoretically does not result from the 
involvement of two $cis$ and $trans$ HNCO isomers in the dissociation dynamics. Rather, we have found that the bimodality is the consequence of an 
impulsive-deflective mechanism suggested by the two trajectories displayed in Fig.~\ref{figure2} (representing more than 80 \% of the 
trajectories). The red path undergoes a strong impulsion from the Wigner region down to the bottom of the well, followed by a deceleration 
insufficient to prevent the system from crossing the $trans$ TS with a large angular velocity $\dot{\alpha}$ and thus, a large rotational 
excitation of CO. The blue trajectory performs one loop before leaving the well, but its path beyond the inner (last) turning point is fairly  
similar to the red path. If all the trajectories were like the two previous ones, the rotational distribution would be a relatively narrow 
unimodal peak centred at $j_{CO} \approx 35$. However, part of the trajectories appear to be deflected by the potential wall lying on the 
right-side of the mountain pass en route to the product valley (see Fig.~\ref{figure2}; this region is approximately defined by $r_{NC}$ within [3.8, 4.8] and 
$\alpha$ below 90 \degree). This is, for instance, slightly the case of the red path. This deflection, if strong enough, may make the 
final direction of the trajectories nearly parallel to $r_{NC}$, leading thereby to a rotationally cold CO. Consequently, the strong 
initial impulsion and the subsequent deflection create the large peak and the smaller bump, respectively, as is clearly visible in the lower 
panels of Fig.~\ref{figure4}.

In summary, the theoretical dynamics of the photofragmentation HNCO $\longrightarrow$ NH(a$^1\Delta$) + CO(X$^1\Sigma$) at 201 and 210 nm 
have been examined, assuming this process entirely takes place in the S$_1$ state. Specifically, accurate semiclassical trajectory calculations on 
a new 6D-PES have been carried out. The pair-correlated kinetic energy distributions (KEDs) and anisotropy parameters (APs) measured by Yang and 
co-workers \cite{Wang2007,PY}, following state-selective detection of either CO or $^1$NH via REMPI, have been calculated.
The positions, widths and shapes of the KEDs are in good agreement with the experimental ones for most CO or $^1$NH states probed.
The APs are also in satisfying agreement with experiment, though the perpendicular polarization of the angular distributions 
is slightly overestimated by theory. On the other hand, we found that the vibrational state distributions (VSDs) at 201 nm do not involve 
any inversion, in contrast with the VSDs deduced from the experimental KEDs \cite{PY}. However, experimentally deduced VSDs are 
necessarily equivocal 
when the co-product state-resolved components of the KEDs overlap, as is the case at 201 nm, whereas the theoretical determination of the VSDs 
is unambiguous. We thus see no contradiction between our results and the measurements of  Yang and co-workers \cite{PY}.
Last but not least, theory precludes the possibility of any link between the existence of two $cis$ and $trans$ HNCO isomers and 
the bimodality observed in the KEDs, as speculated in Ref.\cite{PY}. Instead, our treatment shows that bimodality should be the consequence 
of an impulsive-deflective mechanism due to two repulsive walls of the 6D-PES, the first one producing rotationally hot CO products, 
the second one cooling part of them. These findings provide new insights into the dissociation dynamics of the title tetratomic 
prototype fragmentation.

At the moment, pair-correlated KEDs, VSPs and APs for dissociations involving more than three atoms have been much more often measured than 
theoretically predicted. In this regard, the present work fills an important gap ensuring the synergy between experiment and theory for these
complex processes. This work can be straightforwardly extended to the description of state-to-state photodynamics through imaging techniques
for polyatomics playing crucial roles in atmospheric and environmental chemistry.

\begin{acknowledgments} 
We would like to thank Dr. Gustavo A. Garcia (Synchrotron Soleil, France) for fruitful discussions.
\end{acknowledgments}


\begin{thebibliography}{10}
\section*{References}

\bibitem{Liebig1830}
J. Liebig and F. W\"{o}hler, 
Ann. Phys. \textbf{96}, 369 (1830).
\bibitem{Marcelino2010}
N. Marcelino, S. Br\"{u}nken, J. Cernicharo, D. Quan, E. Roueff, E. Herbst, and P. Thaddeus, 
A\&A \textbf{516}, A105 (2010). 
\bibitem{Perry1986}
R. A. Perry and D. L. Siebers, 
Nature \textbf{324}, 657 (1986).
\bibitem{Miller1991}
 J. A. Miller and C. T. Bowman, 
 Int. J. Chem. Kinet. \textbf{23}, 289 (1991).
\bibitem{Snyder1972}
L. E. Snyder and D. Buhl, 
Astrophys. J. \textbf{177}, 619 (1972).
\bibitem{Buhl1972}
D. Buhl, L. E. Snyder, and J. Edrich, 
Astrophys. J. \textbf{177}, 625 (1972).
\bibitem{Zinchenko2000}
I. Zinchenko, C. Henkel, and  R. Q. Mao, 
A\&A \textbf{361}, 1079 (2000).
\bibitem{Nguyen1991}
Nguyen-Q-Rieu, C. Henkel, J. M. Jackson, and R. Mauersberger, 
A\&A \textbf{241}, L33 (1991).
\bibitem{Jackson1984}
J. M. Jackson, J. T. Armstrong, and A. H. Barrett, 
ApJ \textbf{280}, 608  (1984).
\bibitem{Roberts2011}
J. M. Roberts, P. R. Veres, A. K. Cochran, C. Warneke, I. R. Burling, R. J. Yokelson, B. Lerner, J. B. Gilman, W. C. Kuster, 
R. Fall, and J. de Gouw, PNAS \textbf{108}, 8966 (2011).
\bibitem{Bowman1992}
C. T. Bowman, 
24th Symp. (Int. ) Combust., 859 (1992).
\bibitem{Young2012}
P. J. Young,  L. K. Emmons, J. M. Roberts, J.-F. Lamarque, C. Wiedinmyer, P. Veres, and T. C. VandenBoer, 
J. Geophys. Res. \textbf{117}, D10308 (2012).

\bibitem{Zyrianov}
M. Zyrianov, Th. Droz-Georget, and H. Reisler,                                               
J. Chem. Phys. \textbf{110}, 2059 (1999).

\bibitem{Klo}
J.-J. Klossika and R. Schinke,                                               
 J. Chem. Phys. \textbf{111}, 5882 (1999).
\bibitem{Laine2001}
H. Laine Berghout, Shizuka Hsieh, and F. Fleming Crim, 
J. Chem. Phys. \textbf{114}, 10835 (2001).
\bibitem{Spiglanin1987}
T. A. Spiglanin, R. A. Perry, and D. W. Chandler, J. Chem. Phys. \textbf{87}, 1568 (1987).
\bibitem{Spiglanin1987_2}
T. A. Spiglanin and D. W. Chandler, J. Chem. Phys. \textbf{87}, 1577 (1987). 
\bibitem{Yi1996}
W. K. Yi and R. Bersohn, Chem. Phys. Lett. \textbf{206}, 365 (1993).
\bibitem{Brownsword1996}
R. A. Brownsword, T. Laurent, R. K. Vatsa, H. R. Volpp, and J. Wolfrum, 
Chem. Phys. Lett. \textbf{249}, 162 (1996).
\bibitem{Brownsword1996_2}
R. A. Brownsword, T. Laurent, R. K. Vatsa, H. R. Volpp, and J. Wolfrum, Chem. Phys. Lett. \textbf{258}, 164 (1996).
\bibitem{Brownsword1997}
R. A. Brownsword, M. Hillenkamp, T. Laurent, R. K. Vatsa, and H. R. Volpp, J. Chem. Phys. \textbf{106}, 4436 (1997).
\bibitem{Brownsword1999}
R. A. Brownsword, M. Hillenkamp, H. R. Volpp, and R. K. Vatsa, Res. Chem. Intermed. \textbf{25}, 339 (1999).

\bibitem{Yarkony2016}
D. R. Yarkony, 
J. Chem. Phys. \textbf{114}, 2614 (2016).
\bibitem{Yarkony2001}
D. R. Yarkony, 
Mol. Phys. \textbf{99}, 1463 (2001).
\bibitem{Wang2007}
H. Wang, S.-L. Liu, J. Liu, F.-Y. Wang, B. Jiang, and X.-M. Yang, 
Chinese J. Chem. Phys. \textbf{20}, 388 (2007).
\bibitem{PY}
Z. Zhang, Z. Chen, C. Huang, Y. Chen, D. Dai, D. H. Parker, and X. Yang,
J. Phys. Chem. A \textbf{118}, 2413 (2014).



\bibitem{Xie2014}
C. Xie, J. Ma, X. Zhu, D. H. Zhang, 
D. R. Yarkony, D. Xie, and H. Guo, 
J. Phys. Chem. Lett. \textbf{5}, 1055 (2014).


\bibitem{Part3}
L. Bonnet, 
Int. Rev. Phys. Chem. \textbf{32}, 171 (2013).

\bibitem{Chinese-C+H2}
Z. Shen, H. Ma, C. Zhang, M Fu, Y. Wu, W. Bian, and J. Cao,
Nat. Commun. \textbf{8}, 14094 (2017).

\bibitem{Goursaud}
S. Goursaud, M. Sizun and F. Fiquet-Fayard,
J. Chem. Phys. \textbf{65}, 5453 (1976).

\bibitem{Schinke}
R. Schinke,
Photodissociation Dynamics (Cambridge University Press, Cambridge, 1993). 




\bibitem{Bohr}
N. Bohr,
Phil. Mag. \textbf{26}, 1 (1913).




\bibitem{Miller}
W. H. Miller,                                               
J. Chem. Phys. \textbf{53}, 3578 (1970).

\bibitem{Marcus}
R. Marcus,                                               
 J. Chem. Phys. \textbf{54}, 3965 (1971).

\bibitem{BCE}
L. Bonnet, J. C. Corchado and J. Espinosa-Garcia,
C. R. Chimie \textbf{19}, 571 (2016).






\end{thebibliography}
\end{document}